# Construction and Visualization of Optimal Confidence Sets for Frequentist Distributional Forecasts[*]

David Harris[†], Gael M. Martin[‡], Indeewara Perera[§] and D.S. Poskitt[¶]

August 3, 2017

**Abstract**

The focus of this paper is on the quantification of sampling variation in frequentist probabilistic forecasts. We propose a method of constructing confidence sets that respects the functional nature of the forecast distribution, and use animated graphics to visualize the impact of parameter uncertainty on the location, dispersion and shape of the distribution. The confidence sets are derived via the inversion of a Wald test and are asymptotically uniformly most accurate and, hence, optimal in this sense. A wide range of linear and non-linear time series models - encompassing long memory, state space and mixture specifications - is used to demonstrate the procedure, based on artificially generated data. An empirical example in which distributional forecasts of both financial returns and its stochastic volatility are produced is then used to illustrate the practical importance of accommodating sampling variation in the manner proposed.

---

[*]This research has been supported by Australian Research Council (ARC) Discovery Grants DP120102344 and DP150101728.
[†]Department of Economics, University of Melbourne, Australia.
[‡]Department of Econometrics and Business Statistics, Monash University, Melbourne, Australia. Corresponding author; email: gael.martin@monash.edu.
[§]Department of Econometrics and Business Statistics, Monash University, Melbourne, Australia.
[¶]Department of Econometrics and Business Statistics, Monash University, Melbourne, Australia.

# 1 Introduction

Probabilistic (or distributional) forecasting - namely, the assignment of a probability distribution to the future values of a random variable - is a suitable way of approaching the act of prediction, fitting naturally as it does with the human propensity to quantify uncertainty using probability and to frame forecasts of an uncertain future in probabilistic terms. Probabilistic forecasts are also consistent with the sample space of the variable in question, as well as being replete with all important distributional (in particular tail) information. In contrast, point forecasts, based on single summary measures of central location, convey no such distributional information and, potentially, also lack coherence with the sample space as, for example, when a conditional mean forecast of a discrete random variable assumes real values.

Despite earlier attempts to draw attention to the worth of probabilistic forecasts (e.g. Dawid, 1984), such forecasts have only started to gain some prominence in the literature in more recent times. A focal point of much of this work has been the development of techniques for *ex-post* evaluation of distributional forecasts using observed outcomes. Calibration with realized values is assessed via the probability integral transform method (e.g. Dawid; Diebold *et al.*, 1998; Geweke and Amisano, 2010), predictive accuracy tests (e.g. Corradi and Swanston, 2006; Amisano and Giacomini, 2007), or via the application of calibration criteria in combination with measures of predictive 'sharpness', including the use of various scoring rules (e.g. Gneiting *et al.*, 2007; Gneiting and Raftery, 2007; Czado *et al.*, 2009; Gneiting and Katzfuss, 2014). In contrast, McCabe, Martin and Harris (2011) present the concept of an *ex-ante* efficient estimator of a forecast distribution, within a particular class of (discrete) time series models. Explicit acknowledgement therein of the dependence of the (estimated) forecast distribution on the frequentist properties of the underlying parameter estimates prompted the use of a subsampling method (Politis *et al.*, 1999) to capture sampling variation in a manner that respected the functional nature of the forecast distribution. No attempt was made, however, to extend the procedure beyond the specific model class at hand, to provide optimality results, or to establish a general method of visualization.

The focus of this paper in on the derivation of an optimal method for measuring sampling variation in frequentist distributional forecasts, in *any* time series setting, and the provision of a computational technique for visualizing that variation. Whilst the principle that underpins the approach is completely general, we demonstrate it solely within the context of distributional forecasts produced via *parametric* time series models. In brief, we produce the range of forecast distributions that bound the 'set' or 'region' within which the true forecast distribution lies with a given level of confidence. This range of distributions is, in turn, determined by the range of values for the unknown (possibly vector-valued) parameter, $\theta$ say, that defines the boundary of the confidence set for the given nominal value, with this boundary produced via the inversion of a Wald test. Given that the latter is an asymptotically uniformly most powerful invariant (AUMPI) test in the settings we consider, the arguments of Cox and Hinkley (1974), Le Cam and Yang (1990), Choi, Hall and Schick (1996) and Shao (2003) can be invoked to establish that the confidence sets so produced are asymptotically uniformly most accurate (AUMA). Whilst there would appear to be no unique characterization of the AUMA property, Cox and Hinkley (Section 7.2) highlight the fact that a uniformly most powerful test produces a confidence set that includes a false value of $\theta$ with minimum probability, and which is optimal in that particular sense. Moreover, they point out that in certain simple (scalar) cases, with a pivotal test statistic, this optimality translates into the confidence interval being the narrowest such interval on the real line. Shao (2003, Theorem 7.6) extends this link to the multivariate case, making the comparable connection between the confidence region with the smallest volume and the minimum coverage probability (of false values). Drawing on this form of motivation and, importantly, recognizing the computational advantages lent by the quadratic form of the Wald statistic, we pursue the production of optimal confidence sets for distributional forecasts solely via the Wald test route.

We make it explicit from the outset that the exercise is undertaken with the frequentist forecasting paradigm in mind. In this case the substitution of 'plug-in' estimators of unknown parameters into a forecast distribution renders the latter a random function with sampling variation that reflects parameter uncertainty. It is this sampling variation that we are attempting to capture, and in a way that respects the integration to unity property of the random function of interest. This situation contrasts with that which prevails under the Bayesian paradigm, in which forecast distributions condition solely on past data, with unknown parameters integrated out via the Bayesian probability calculus. The impact of parameter uncertainty in that case is directly reflected in the form of the (single) forecast distribution so produced, with no additional measurement step required. (See, for example, Geweke, 2005, for a textbook treatment of Bayesian forecasting).

We also emphasize the contrast between the approach adopted in this paper and other methods for representing (frequentist) sampling variation in a forecasting context. Such methods have focussed on: the production of *point-wise* confidence intervals for estimated forecast prob-abilities (Freeland and McCabe, 2004); the use of the bootstrap to extend (marginal) prediction intervals to cater for parameter uncertainty (see De Gooijer and Hyndman, 2006, for an extensive survey; and Rodriguez and Ruiz, 2009, 2012, for recent applications); or the use of the bootstrap to construct joint prediction regions (over multiple forecast horizons) with correct (asymptotic) coverage in the presence of estimation error (Wolf and Wunderli, 2015). *Our* focus, we reiterate, is on representing the impact of sampling variation on the *full* forecast distribution for any single forecast horizon, and visualizing the way in which that variation influences all aspects of that distribution: location, dispersion and shape.

The outline of the paper is as follows. In Section 2 we define the problem at hand, namely the production of an estimated forecast distribution and the recognition of the random nature of that quantity. We outline the approach that we adopt in producing forecast distributions that define the 'boundary' of the confidence set over which sampling variation occurs, based on the inversion of a Wald test procedure, with both unconditional and conditional versions of the test entertained. The numerical technique used to compute that boundary is described. In Section 3 a range of time series models, that encompass long memory, state space and mixture models, are used to illustrate the proposed ideas. We highlight the wide range of forecast distributions that *could* be encountered in hypothetical repeated sampling (even in these simple examples) and provide visualization of that distributional variation using animated graphics. An empirical illustration using returns and (an observable measure of) volatility of the S&P500 stock index is provided in Section 4. As part of that illustration we demonstrate the implications of parameter variation for conclusions drawn regarding the predictive superiority of one model over another. In short, scalar scoring rules (such as the logarithmic and quadratic scores used in the demonstration) reflect the influence of parameter variation, as do the differences between corresponding scores for two alternative models. In particular, parameter variation can induce variation in the *sign* of score differences and, hence, alter the conclusion about relative predictive performance. This effect is illustrated graphically for the particular financial models entertained. Section 5 concludes with discussion of some matters that remain unresolved and that form the basis of ongoing investigations by the authors.

## 2   Frequentist Distributional Forecasts and Confidence Sets

Assume a time series of random variables, $Y_1, Y_2, ..., Y_T$, with observed values collected in the vector $y_{1:T} = (y_1, \ldots, y_T)'$. Without loss of generality, the object of interest is the conditional forecast distribution,

$$f(.|y_{1:T};\theta_0),$$

for $Y_{T+1}$, the unknown random variable at time $T+1$, with $\theta_0$ the true value of the $(p\times 1)$ vector of unknown parameters. We use $f\left(.|y_{1:T};\hat{\theta}(y_{1:T})\right)$ as the estimated (one-step-ahead) forecast distribution, where the average log-likelihood function is given by $\ell_T(\theta) = T^{-1}\sum_{t=1}^{T}\log f(y_t|y_{1:t-1};\theta)$ and

$$\hat{\theta}(y_{1:T}) = \arg\max_{\theta} \ell_T(\theta). \qquad (1)$$

The subscript $T$ is used here as a reminder of the fact that what we refer to as the 'log-likelihood function' without further qualification is indeed the average log-likelihood function over the sample. Viewed as a random *estimator*, $\hat{\theta}(Y_{1:T})$ in (1) has a sampling distribution which, in turn, induces random sampling variation in the function $f(.|y_{1:T};\hat{\theta}(Y_{1:T}))$. For the purposes of this paper we assume that the parametric form of $f(.|y_{1:T};\theta)$ is known, although extension to semi-parametric models could be undertaken without altering the qualitative nature of the key points made herein. Extension to the case in which $f(Y_{T+k}|y_{1:T};\theta_0)$ is the object of interest, with $k > 1$, is also conceptually straightforward and, hence, not considered.

Throughout the paper we adopt the conventional approach of defining the forecast distribution as the quantity in which the conditioning values, $y_{1:T}$, are fixed at the observed values.[1] Estimation of $f(.|y_{1:T};\theta_0)$ is thus viewed as a functional estimation problem in terms of $Y_{T+1}$ for a fixed $y_{1:T}$. The point estimator of $f(.|y_{1:T};\theta_0)$ is $f(.|y_{1:T};\hat{\theta}(Y_{1:T}))$ and an asymptotically valid confidence set for $f(.|y_{1:T};\theta_0)$ is a set $\mathcal{F}_\alpha(Y_{1:T})$ of distributions such that

$$\Pr(f(.|y_{1:T};\theta_0) \in \mathcal{F}_\alpha(Y_{1:T})) \underset{T\to\infty}{\to} 1-\alpha.$$

As noted in the Introduction a confidence set is considered 'good' if the 'size' of $\mathcal{F}_\alpha(.)$ is 'small', somehow defined, and we seek the optimal confidence set in this sense via the inversion of an AUMPI test. With the Wald procedure adopted as the underlying test, the quadratic form of the test statistic allows the inversion to occur via an ellipsoid, as described in Section 2.2. We are also interested in displaying graphically distributions on the boundary of $\mathcal{F}_\alpha(.)$, in order to illustrate visually the range of forecast distributions that could possibly occur, in hypothetical repeated sampling, in a range of different models.

---

[1] This approach could be re-phrased as one in which $\hat{\theta}_T(Y_{1:T})$ is assumed to be independent of the conditioning values, in which case sampling variation in $f\left(.|y_{1:T};\hat{\theta}(Y_{1:T})\right)$ can be viewed as being driven by the (marginal) sampling distribution of $\hat{\theta}(Y_{1:T})$, rather than by the distribution of $\hat{\theta}(Y_{1:T})$ conditional on any observed conditioning values that define $f\left(.|y_{1:T};\hat{\theta}(Y_{1:T})\right)$; see Phillips (1979) for some early discussion of related issues. We return briefly to this issue of conditioning in the Discussion section that completes the paper.

## 2.1 Optimal confidence sets based on the inversion of Wald tests

The standard Wald test statistic for $H_0: \theta_0 = \theta$ against $H_1: \theta_0 \neq \theta$ is

$$\omega(\theta) = T\left(\hat{\theta} - \theta\right)' V(\hat{\theta})^{-1} \left(\hat{\theta} - \theta\right), \tag{2}$$

where $V(\hat{\theta})^{-1}$ is a consistent estimator of $i(\theta_0) = -\lim_T T^{-1} \sum_{t=1}^{T} E[h_t(\theta_0)]$ and $h_t(\theta) = \frac{\partial^2 \ln f(y_t|y_{1:t-1};\theta)}{\partial\theta\partial\theta'}$. (Note that for the sake of notational simplicity, in (2) and hereinafter we denote the estimator of $\theta_0$ as $\hat{\theta}$, rather than as $\hat{\theta}(Y_{1:T})$.)

Under the null, $\omega(\theta) \xrightarrow{P} \chi^2(p)$ and, as proven initially in Wald (1943) and is subsequently standard knowledge, the test is AUMPI under regularity. Now defining $c_\alpha$ as the $\alpha$-level critical value from the (asymptotically valid) $\chi^2(p)$ distribution, a $(1-\alpha)100\%$ confidence set is defined as the set of null values not rejected:

$$\mathcal{C}_\alpha^\omega = \{\theta : \omega(\theta) \leq c_\alpha\}. \tag{3}$$

By the definition of $c_\alpha$ this set has coverage property, $\mathrm{pr}(\theta_0 \in \mathcal{C}_\alpha^\omega) \xrightarrow[T \to \infty]{} 1 - \alpha$. We then define the forecast confidence set for the true forecast distribution, $f(.|y_{1:T};\theta_0)$, as

$$\mathcal{F}_\alpha(.) = \{f(.|y_{1:T};\theta) : \theta \in \mathcal{C}_\alpha^\omega\} \tag{4}$$

which, by definition, has the same coverage property as $\mathcal{C}_\alpha^\omega$. By the arguments cited earlier, the confidence set in (3) (equivalently that in (4)) is AUMA and, hence, viewed as optimal.

The forecast distributions on the boundary of $\mathcal{F}_\alpha(.)$ are, by construction, characterized by values of $\theta$ on the boundary of the $(1-\alpha)100\%$ confidence set for the parameters. Depending on the nature of the problem, and the role of $\theta$ therein, the nature of these forecast distributions can differ substantially, one from the other, and one aim of the paper is to highlight that fact by visualizing the distributions that can arise, at the extreme end of the spectrum, in a variety of models. For one simple example, we also suggest a way of selecting 'representative' extreme distributions, thereby avoiding the need to represent the full range of possibilities pictorially. Given the requirement to solve (3) for $\theta$ in order to define (4), the algebraic form of the test statistic, and the interpretation of the parameters themselves, obviously plays a role in the selection of such representative distributions on the boundary.

In view of the quadratic form of (2), the boundary set in (3) is an ellipsoid. In Section 2.2 we describe a numerical method to traverse the surface of such a boundary set. As is made clear therein, the computational burden increases with the dimension of the parameter set. Hence, we suggest a dimension reduction technique that may be appropriate in some cases. Specifically, it is helpful to express the forecast distribution in terms of its 'canonical' or fundamental parameters; for example, the (conditional) mean and variance for a Gaussian linear model. Denoting this (vector) parameter as $\eta(\theta; y_{1:T})$, the Wald test statistic for $H_0 : \eta(\theta_0; y_{1:T}) = \eta(\theta; y_{1:T})$ against $H_1 : \eta(\theta_0; y_{1:T}) \neq \eta(\theta; y_{1:T})$ is thus $T\{\eta(\hat{\theta}; y_{1:T}) - \eta(\theta; y_{1:T})\}'\Sigma(\hat{\theta})^{-1}\{\eta(\hat{\theta}; y_{1:T}) - \eta(\theta; y_{1:T})\}$, where: $\Sigma = \nabla(\hat{\theta}; y_{1:T})V(\hat{\theta})\nabla(\hat{\theta}; y_{1:T})'$, with $\nabla(\theta; y_{1:T}) = \partial\eta(\theta; y_{1:T})/\partial\theta'$, and if $\eta(\hat{\theta}; y_{1:T})$ is a $p$-dimensional vector, then the $(1-\alpha)100\%$ confidence set for $\eta(\hat{\theta}; y_{1:T})$ is given by

$$\left\{\eta(\theta; y_{1:T}) : \omega(\eta(\theta; y_{1:T})) := T[\eta(\hat{\theta}; y_{1:T}) - \eta(\theta; y_{1:T})]'\Sigma(\hat{\theta})^{-1}[\eta(\hat{\theta}; y_{1:T}) - \eta(\theta; y_{1:T})] \leq c_\alpha\right\}. \quad (5)$$

Henceforth the confidence set in (5) is referred to as the 'conditional confidence set', in contrast to the (unconditional) confidence set in (3). In the simplest case in which the problem is described by a scalar conditional mean and conditional variance, the quadratic form in (5) describes a two-dimensional mean/variance ellipse in $\eta(\theta; y_{1:T})$, which is all that is necessary to describe the confidence set for $f(.|\eta(y_{1:T}; \theta))$. This dimension reduction technique is demonstrated in Section 3.2 and exploited in the empirical illustration in Section 4.

## 2.2 Numerical solution for the bounding set

In this section we describe the numerical method used to traverse the boundary of the confidence set for the forecast distribution. Without loss of generality we use the four-dimensional case for illustration, selecting a grid of points on the bounding set in (3) associated with the unconditional Wald test. The method is clearly applicable to the bounding values in the conditional confidence set in (5) as well, in which case the initial dimensionality of the problem would be reduced accordingly.

Let $\mathcal{B}_\alpha = \{T(\hat{\theta} - \theta)'V(\hat{\theta})^{-1}(\hat{\theta} - \theta) = c_\alpha\}$ denote the $(1-\alpha)100\%$ bounding set for the four-dimensional vector $\theta$ obtained by inverting the Wald test, and let $U$ be a matrix square root of $V(\hat{\theta})^{-1}$ such that $V(\hat{\theta})^{-1} = UU'$, produced, for example, via a Cholesky decomposition of $V(\hat{\theta})^{-1}$. Then, $T(\hat{\theta} - \theta)'V(\hat{\theta})^{-1}(\hat{\theta} - \theta) = c_\alpha$ may be written as $(\hat{\theta} - \theta)'\frac{UU'}{c_\alpha/T}(\hat{\theta} - \theta) = 1$, and hence, with $x = U'(\hat{\theta} - \theta)/\sqrt{c_\alpha/T}$, as $x'x = 1$. Therefore, if one can traverse the unit hyper-sphere defined by $x'x = 1$ then for each $x$ the corresponding $\theta$ may be recovered as $\theta = \hat{\theta} - \sqrt{c_\alpha/T}[U^{-1}]'x$.

Traversing $x'x = 1$ may be undertaken in several different ways. For example, and with reference to the four dimensional case, one may use the polar coordinates: $x_1 = \cos(\lambda_1)$, $x_2 = \sin(\lambda_1)\cos(\lambda_2)$, $x_3 = \sin(\lambda_1)\sin(\lambda_2)\cos(\lambda_3)$ and $x_4 = \sin(\lambda_1)\sin(\lambda_2)\sin(\lambda_3)\cos(\lambda_4)$, for $\lambda_1, \lambda_2, \lambda_3 \in [0, \pi]$ and $\lambda_4 \in [0, 2\pi)$. This represents a fairly natural way of constructing grids in the four angles to traverse the sphere, and for each point to be translated back to $\theta$ and, hence the forecast distribution, as a function of $\theta$. Clearly this generalizes (in principle) to any number of dimensions; however, as is the case with any grid-based deterministic method, the so-called 'curse of dimensionality' applies, with the computational burden being exponential in the number of dimensions. It is for this reason that the dimension reduction afforded by the conditional method can yield benefits, with conditional method involving the replacement of $\hat{\theta}$, $\theta$ and $V(\hat{\theta})^{-1}$ by $\eta(\hat{\theta}; y_{1:T}), \eta(\theta; y_{1:T})$ and $\Sigma(\hat{\theta})^{-1}$, respectively.

In the next section we use this numerical technique to produce the bounding distributions for the forecast distributions associated with several different time series models. In the first example the parameter space is two dimensional, in which case the bounding set for $\theta$ is an ellipse and it is easy to extract and display 'representative' bounding forecast distributions. In addition, an animated plot can be used to display the full range of bounding distributions associated with traversing (discretely) the ellipse of values for $\theta$. In the second example the problem is multi-dimensional, and encompasses a wide range of time series models, including long memory models and state space specifications; however the conditioning method described in the previous section can be used to reduce the problem to a two-dimensional one, rendering description (and visualization) of sampling variation in the forecast distribution straightforward. In both cases both the location and variance of the forecast distribution are affected by the variation in $\theta$. In the final two examples, which are three-dimensional and four-dimensional respectively, $\theta$ impacts on the higher-order moments of the forecast distribution, with there being a much wider range of possible distributions at the $(1-\alpha)100\%$ boundary as a consequence. For the purpose of illustration, we set $1-\alpha = 0.95$ in all examples.

## 3 Illustrations

### 3.1 Gaussian AR(1)

We begin by considering a simple two parameter problem, that of a stationary Gaussian autoregressive model of order one (AR(1)):

$$Y_t = \alpha_1 Y_{t-1} + U_t, \tag{6}$$

where $U_t \sim i.i.d.N(0, \sigma^2)$, with $\sigma^2 > 0$ and $|\alpha_1| < 1$. With reference to the general notation defined above, we have $\theta = (\alpha_1, \sigma^2)'$. Given (6), the (average) log-likelihood function is $\ell_T(\theta) = \left[-\frac{1}{2}\log 2\pi - \frac{1}{2}\log \sigma^2 - \frac{1}{2T\sigma^2}\sum_{t=1}^{T}(y_t - \alpha_1 y_{t-1})^2\right]$, conditional on $y_0$, and the (condi-tional) MLEs of the elements of $\theta$ have the familiar form, $\widehat{\alpha}_1 = \left(\sum_{t=1}^T y_t y_{t-1}\right) / \left(\sum_{t=1}^T y_{t-1}^2\right)$ and $\widehat{\sigma}^2 = T^{-1} \sum_{t=1}^T (y_t - \widehat{\alpha}_1 y_{t-1})^2$. The Wald test statistic for $H_0 : (\alpha_{10}, \sigma_0^2)' = (\alpha_1, \sigma)'$ against $H_1 : (\alpha_{10}, \sigma_0^2)' \neq (\alpha_1, \sigma)'$ assumes the following form:

$$\omega(\theta) = \frac{T}{2} \left(\frac{\sigma^2}{\widehat{\sigma}^2} - 1\right)^2 + \frac{(\widehat{\alpha}_1 - \alpha_1)^2}{\sigma^2} \sum_{t=1}^T y_{t-1}^2,$$

and the $(1-\alpha)100\%$ confidence set for $(\alpha_{10}, \sigma_0^2)$ in (3) defined accordingly, with $c_\alpha$ the $(1-\alpha)100\%$ quantile of the $\chi^2(2)$ distribution. This defines a two-dimensional region bounded by an ellipse centered at $(\alpha_1, \sigma^2) = (\widehat{\alpha}_1, \widehat{\sigma}^2)$. Alternatively, using a transformation of variables, $X = \left(\frac{\widehat{\sigma}^2}{\sigma^2} - 1\right)$, $Y = \frac{(\widehat{\alpha}_1 - \alpha_1)}{\sigma}$, we can consider the ellipse defined by

$$\bar{\mathcal{T}} = \left\{(X,Y) : \frac{X^2}{a^2} + \frac{Y^2}{b^2} = 1\right\}, \tag{7}$$

where

$$a := \left\{\frac{c_\alpha}{(T/2)}\right\}^{1/2}, \quad b := \left\{\frac{c_\alpha}{\sum_{t=1}^T y_{t-1}^2}\right\}^{1/2}. \tag{8}$$

The set $\bar{\mathcal{T}}$ has a one-to-one correspondence with the ellipse bounding (3), with the center $(X,Y) = (0,0)$ corresponding to $(\alpha_1, \sigma^2) = (\widehat{\alpha}_1, \widehat{\sigma}^2)$. The bounding ellipse in (3) or, equivalently, in (7), thus defines an infinite number of pairs of values for $(\alpha_1, \sigma^2)$ that can be defined as 'extreme' and that can be extracted to define forecast distributions that bound the forecast confidence set in (4) via the numerical method described in Section 2.2.

In Figure 1 the forecast distributions defined by traversing the boundary in (7), with $1 - \alpha = 0.95$, are reproduced, in animation, for $T = 100$ and $(\alpha_{10}, \sigma_0^2) = (0.6, 1)$. The single empirical forecast distribution is superimposed - represented by the (fixed) dotted line. The full extent of the variation - in both location and dispersion - of the forecast distributions on the $95th$ percentile, is in evidence, alerting the investigator to the varied probabilistic statements about the unknown $Y_{T+1}$ that *could* arise due to parameter uncertainty.

Whilst the dynamic display in Figure 1 is instructive, in this simple two-parameter case it is also possible to identify *representative* pairs of bounding distributions, based on simultaneously choosing the pairs of largest and smallest of values on the ellipse for: 1) the conditional forecast variance, $\sigma^2$; and 2) the parameter $\alpha_1$, and, equivalently, the conditional forecast mean, $\alpha_1 y_T$, using the following steps:

1. Firstly, as $-a \leq \widehat{\sigma}^2/\sigma^2 - 1 \leq a$, with $a$ as defined in (8), then $\frac{\widehat{\sigma}^2}{1+a} \leq \sigma^2 \leq \frac{\widehat{\sigma}^2}{1-a}$. The bounding forecast distribution with the largest (resp. smallest) conditional variance thus

Figure 1: Forecast distributions at the 95th percentile boundary of the confidence set in (4), for data generated artificially from (6), with $(\alpha_{10}, \sigma_0^2) = (0.6, 1)$ and $T = 100$. The solid lines represent the bounding distributions, whilst the empirical estimate of the true distribution is given by the dotted line.

corresponds to the boundary point:

$$(\alpha_1, \sigma^2) = \left(\widehat{\alpha}_1, \widehat{\sigma}^2/(1-a)\right) \text{ (resp. } (\alpha_1, \sigma^2) = \left(\widehat{\alpha}_1, \widehat{\sigma}^2/(1+a)\right) \text{ )}. \tag{9}$$

2. Next, if $(\alpha_1, \sigma^2)$ lies on the boundary of $\mathcal{C}_\alpha^\omega$, then for any given $\sigma^2$,

$$\alpha_1 = \widehat{\alpha}_1 \pm \sqrt{\frac{\sigma^2}{\sum_{t=1}^T y_{t-1}^2} \left\{ c_\alpha - \frac{T}{2}\left(\frac{\widehat{\sigma}^2}{\sigma^2} - 1\right)^2 \right\}}.$$

The value of $\sigma^2$ that produces the bounding values of $\alpha_1$ is thus

$$\sigma_m^2 := \arg\max_{\left(\frac{\widehat{\sigma}^2}{1+a} \leq \sigma^2 \leq \frac{\widehat{\sigma}^2}{1-a}\right)} \left\{ c_\alpha - \frac{T}{2}\left(\frac{\widehat{\sigma}^2}{\sigma^2} - 1\right)^2 \right\} \sigma^2$$

and the bounding values for $\alpha_1$ are respectively

$$\alpha_1^U = \widehat{\alpha}_1 + \sqrt{\frac{\sigma_m^2}{\sum_{t=1}^T y_{t-1}^2} \left\{ c_\alpha^* - \frac{T}{2}\left(\frac{\widehat{\sigma}^2}{\sigma_m^2} - 1\right)^2 \right\}}$$

and

$$\alpha_1^L = \widehat{\alpha}_1 - \sqrt{\frac{\sigma_m^2}{\sum_{t=1}^T y_{t-1}^2} \left\{ c_\alpha^* - \frac{T}{2}\left(\frac{\widehat{\sigma}^2}{\sigma_m^2} - 1\right)^2 \right\}}.$$

The bounding forecast distribution with the largest (resp. smallest) degree of persistence thus corresponds to the boundary point:

$$(\alpha_1, \sigma^2) = \left(\alpha_1^U, \sigma_m^2\right) \text{ (resp. } (\alpha_1, \sigma^2) = \left(\alpha_1^L, \sigma_m^2\right) \text{ )}. \tag{10}$$

Clearly, given the one-to-one relationship between $\alpha_1$ and the conditional forecast mean in this case, the four bounding intervals in (9) and (10) also form the basis for bounding the forecast uncertainty associated with estimation of the conditional mean itself.

In Figure 2 we illustrate the bounds in (9) graphically for a case of a sample size of $T = 100$ generated artificially from (6) with $(\alpha_{10}, \sigma_0^2) = (0.6, 1)$. The corresponding plots associated with the bounds in (10) are given in Figure 3. Each figure reproduces the estimated probability density function (pdf) along with the two bounding pdfs. Such figures provide a visual snapshot of the most extreme outcomes (in both dimensions) that could be observed in hypothetical repeated sampling.

Figure 2: Forecast distributions at the 95th percentile boundary of the confidence set in (4) with the largest and smallest conditional variance. Data is generated artificially from (6), with $(\alpha_{10}, \sigma_0^2) = (0.6, 1)$ and $T = 100$. The dotted line depicts the empirical estimate of the true distribution; the dashed (dash-dot) line the bounding distribution associated with the smallest (largest) conditional variance.

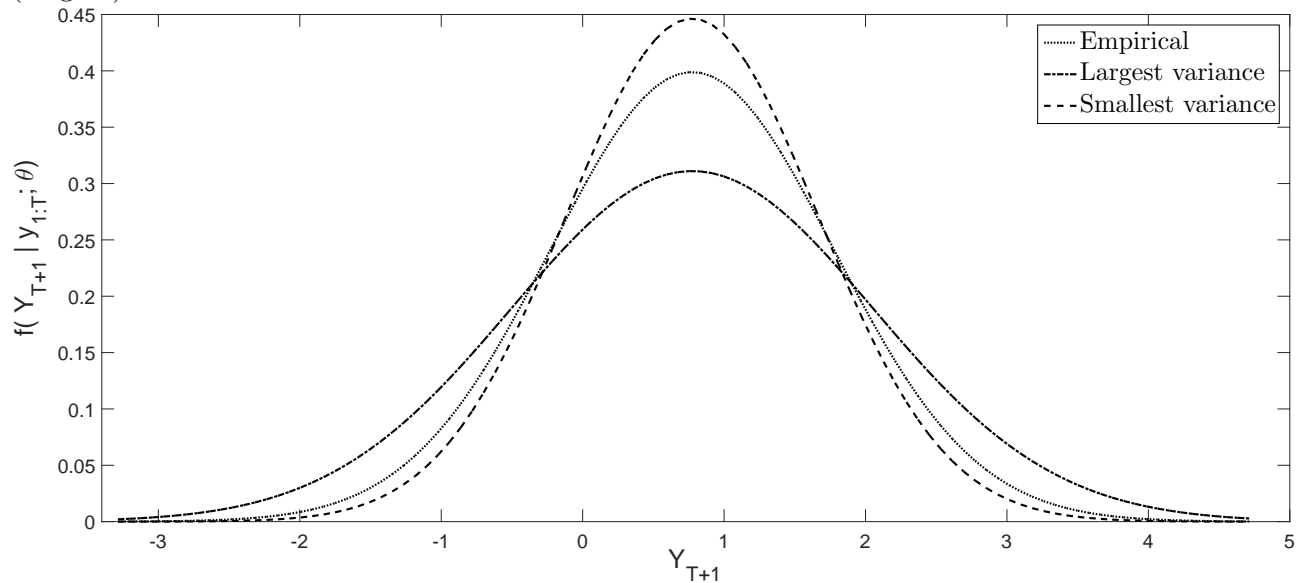

Figure 3: Forecast distributions at the 95th percentile boundary of the confidence set in (4) with the largest and smallest degree of persistence. Data is generated artificially from (6), with $(\alpha_{10}, \sigma_0^2) = (0.6, 1)$ and $T = 100$. The dotted line depicts the empirical estimate of the true distribution; the dashed (dash-dot) line the bounding distribution associated with the smallest (largest) degree of persistence (or conditional mean value).

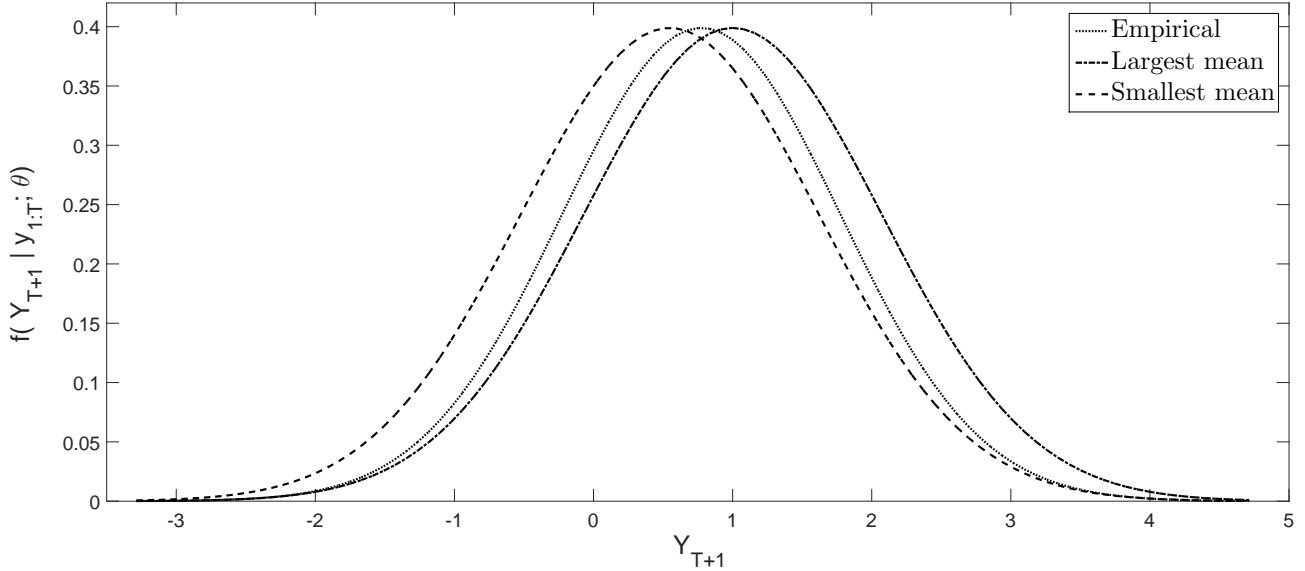

## 3.2 Gaussian linear time series model

Now suppose that

$$Y_t|Y_{1:t-1} \sim N\left(\sum_{j=1}^{t-1} \psi_j(\rho) Y_{t-j},\ \sigma^2\right), \tag{11}$$

where the $\psi_j(\rho)$ may be, for example, the linear process weights defined by a stationary autoregressive (fractionally integrated) moving average AR(FI)MA model with $p \times 1$ parameter vector $\rho$ or, indeed, a linear homoscedastic Gaussian state space model. Defining $x_{t-1}(\rho) = \sum_{j=1}^{t-1} \psi_j(\rho) y_{t-j}$ and $z_{t-1}(\rho) = \left(\sum_{j=1}^{t-1} \{\partial \psi_j(\rho)/\partial \rho\} y_{t-j}\right)$ we can define the average log-likelihood as $\ell_T(\theta) = -\frac{1}{2}\log 2\pi - \frac{1}{2}\log \sigma^2 - \frac{1}{2T\sigma^2} \sum_{t=1}^{T} (y_t - x_{t-1}(\rho))^2$, where $\theta = (\rho', \sigma^2)'$. The unconditional Wald test statistic is thus

$$\omega(\theta) = T \begin{pmatrix} \widehat{\rho} - \rho \\ \widehat{\sigma}^2 - \sigma^2 \end{pmatrix}' \begin{pmatrix} \frac{1}{T\widehat{\sigma}^2} \sum_{t=1}^{T} z_{t-1}(\widehat{\rho}) z_{t-1}(\widehat{\rho})' & 0 \\ 0 & 1/(2\widehat{\sigma}^4) \end{pmatrix} \begin{pmatrix} \widehat{\rho} - \rho \\ \widehat{\sigma}^2 - \sigma^2 \end{pmatrix},$$

with the boundary of (3) defining a $(p+1)-$ dimensional ellipse in $\rho$ and $\sigma^2$, and $c_\alpha$ being the $(1-\alpha)100\%$ quantile of the $\chi^2(p+1)$ distribution.

Clearly, with $p$ potentially very large for some models in the linear class, selecting distinct $(p+1)$-dimensional sets of parameters on the boundary - and their associated forecast distributions - is computationally burdensome. In this case the conditional Wald approach provides a more workable alternative. To this end, we express the forecast mean for $Y_{T+1}$ as $\mu = \sum_{j=1}^{T} \psi_j(\alpha) y_{T+1-j}$ and its estimate as $\hat{\mu} = \sum_{j=1}^{T} \psi_j(\hat{\rho}) y_{T+1-j}$, with $\hat{\rho}$ defined as the MLE of $\alpha$. Given $\partial \mu / \partial \alpha' = z_{T+1}(\alpha)$ and

$$\Upsilon = \begin{pmatrix} \frac{1}{T} z_{T+1}(\hat{\rho}) \left( \sigma^{-2} \sum_{t=1}^{T} z_{t-1}(\hat{\rho}) z_{t-1}(\hat{\rho})' \right)^{-1} z_{T+1}(\hat{\rho})' & 0 \\ 0 & 1/(2\hat{\sigma}^4) \end{pmatrix},$$

it follows that the boundary of

$$\mathcal{C}_\alpha^\omega = \left\{ (\mu, \sigma^2) : T \begin{pmatrix} \hat{\mu} - \mu \\ \hat{\sigma}^2 - \sigma^2 \end{pmatrix}' \Upsilon \begin{pmatrix} \hat{\mu} - \mu \\ \hat{\sigma}^2 - \sigma^2 \end{pmatrix} \leq c_\alpha \right\}$$

defines a two-dimensional ellipse in $\mu$ and $\sigma^2$. In this case, a similar rationale to that adopted for the two-dimensional problem in Section 3.1 could readily be adopted. We reiterate that the approach delineated here, and associated computations, is applicable to any model that can be expressed in the form of (11), highlighting the generality of the method.

### 3.3 Non-Gaussian AR(1)

All examples thus far are characterized by a (conditional) mean and variance component only. When allied with the use of a conditional Wald test, this has amounted to a two-dimensional problem, with the properties of the bounding ellipse able to be exploited, and some meaningful representative bounding forecast distributions extracted. In this section we consider a non-Gaussian example. To simplify the exposition we focus only on the AR(1) case and the unconditional Wald method, noting that more complexity in the conditional mean (and/or variance) could be readily managed via the conditional Wald test described above.

Assume the AR(1) model:

$$Y_t = \alpha_1 Y_{t-1} + U_t, \ U_t \sim F_{skt}, \tag{12}$$

where $F_{skt}$ denotes the skewed Student $t$ distribution of Hansen (1994) with degrees of freedom parameter $v$ and skewness parameter, $\lambda$, and where this particular form of non-Gaussian innovation is chosen for illustrative purposes only. Let $f_{skt}$ denote the pdf of $F_{skt}$,

$$f_{skt}(y; v, \lambda) = \begin{cases} bc \left[ 1 + \frac{1}{v-2} \left( \frac{bx+a}{1-\lambda} \right)^2 \right]^{-\frac{(v+1)}{2}}, & \text{if } x < -\frac{a}{b} \\ bc \left[ 1 + \frac{1}{v-2} \left( \frac{bx+a}{1+\lambda} \right)^2 \right]^{-\frac{(v+1)}{2}}, & \text{if } x \geq -\frac{a}{b}, \end{cases}$$

where $a = 4\lambda c \left( \frac{v-2}{v-1} \right)$, $b = \sqrt{1 + 3\lambda^2 - a^2}$ and $c \frac{\Gamma(\{v+1\}/2)}{\sqrt{\pi(v-2)}\Gamma(v/2)}$, with $v > 2$ and $-1 < \lambda < 1$. Then, the average log-likelihood is

$$\ell_T(\theta) = T^{-1} \sum_{t=1}^{T} \log \left[ f_{skt}(y_t - \alpha_1 y_{t-1}; v, \lambda) \right]; \quad \theta = (\alpha_1, v, \lambda)', \tag{13}$$

and the one-step-ahead forecast distribution is given by $f(Y_{T+1}|y_T; \theta) = f_{skt}(Y_{T+1} - \alpha_1 y_T; v, \lambda)$. Given $\hat{\theta}(y_{1:T}) = \arg\max_\theta \ell_T(\theta)$ one obtains that under $H_0 : \theta_0 = \theta$, $\omega(\theta) \xrightarrow{d} \chi^2(3)$ as $T \to \infty$, with $\omega(\theta)$ as defined in (2), where $V(\hat{\theta})^{-1}$ therein is a consistent estimator of the information matrix, which is unavailable analytically in this case. Denoting the elements in $V(\hat{\theta})^{-1}$ by $V(\hat{\theta})^{-1} = [a_{ij}]$, then values on the boundary of the $(1-\alpha)100\%$ confidence set for $\theta_0$, as given in (3), with $c_\alpha$ the appropriate critical value from the $\chi^2(3)$ distribution, define the surface of a 3-dimensional ellipsoid in $(\alpha_1, v, \lambda)$ centered at the MLEs, $(\hat{\alpha}_1, \hat{v}, \hat{\lambda})$, where the latter are obtained via numerical maximization of (13). Based on data generated from (12), with $T = 100$ and $(\alpha_{10}, v_0, \lambda_0) = (0.8, 5, 0.5)$, we extract values of the triple $(\alpha_1, \nu, \lambda)$ from the ellipsoid in the manner discussed in Section 2.2. Figure 4 provides the associated animated plot of the forecast distributions corresponding to these parameter triples (and for $1 - \alpha = 0.95$). The very wide range of distributional shapes - and associated conclusions regarding the future $Y_{T+1}$ - that could legitimately arise is thereby highlighted.

Figure 4: Forecast distributions at the 95th percentile boundary of the confidence set in (4), for data generated artificially from (12), with $(\alpha_{10}, v_0, \lambda_0) = (0.8, 5, 0.5)$ and $T = 100$. The solid lines represent the bounding distributions, whilst the empirical distribution is given by the dotted line.

## 3.4 Mixture time series model

We complete the set of illustrations by adopting a model in which bimodality in the forecast distribution may feature. To wit, we assume an observable state variable $d_t$ that evolves according to a simple Markov chain with transitions: $\text{pr}(d_t = 1 \mid d_{t-1} = 1) = p_{1|1}$; $\text{pr}(d_t = 1 \mid d_{t-1} = 0) = p_{1|0}$.

The time series of interest $Y_t$ is then $Y_t \mid d_t, \mathcal{H}_{t-1} \sim N\left(d_t\mu_1 + (1-d_t)\mu_0, \sigma^2\right)$, where $\mathcal{H}_{t-1}$ is the information set available at time $t-1$, and $\mu_1, \mu_0, \sigma^2$ are real constants with $\sigma^2 > 0$. In this simple example the information set $\mathcal{H}_{t-1}$ comprises $d_{t-1}$ only. The forecast distribution is either of the following two mixtures of normals

$$Y_{T+1} \mid \{d_T = 1\} \sim p_{1|1} N(\mu_1, \sigma^2) + (1 - p_{1|1}) N(\mu_0, \sigma^2)$$
$$Y_{T+1} \mid \{d_T = 0\} \sim p_{1|0} N(\mu_1, \sigma^2) + (1 - p_{1|0}) N(\mu_0, \sigma^2),$$

where we are making explicit here the dependence of $Y_{T+1}$ on the two discrete values of $d_T$, of which $\mathcal{H}_T$ is comprised. In this case the MLEs of $\mu_1$ and $\mu_0$ are readily available, and the MLE of $\sigma^2$ is produced in the usual way via the regression residuals, $(Y_t - \hat{Y}_t) = Y_t - (\hat{\mu}_1 d_t + \hat{\mu}_0(1 - d_t))$, where $\hat{d}_t = \hat{p}_{1|1} d_{t-1} + \hat{p}_{1|0}(1 - d_{t-1})$. The more common, and empirically relevant, case of $d_t$ unobservable, as well as the case where $\mathcal{H}_{t-1}$ includes lagged values of $Y_t$ (or additional lags of $d_t$) pose additional computational challenges, but provide no additional conceptual insights for the purpose here, and so are not considered.

Now define $\hat{\theta} = (\hat{\mu}_1, \hat{\mu}_0, \hat{\sigma}^2, \hat{p}_1)'$, where $\hat{p}_1 = \hat{p}_{1|1} d_T + \hat{p}_{1|0}(1 - d_T)$. Then, the estimated forecast distribution for $Y_{T+1} \mid \mathcal{H}_T$ is

$$Y_{T+1} \mid \mathcal{H}_T \sim \hat{p}_1 N(\hat{\mu}_1, \hat{\sigma}_T^2) + (1 - \hat{p}_1) N(\hat{\mu}_{0T}, \hat{\sigma}_T^2). \tag{14}$$

Note that this forecast distribution is conditional on (the observed) $d_T$ and this state is held fixed in the confidence interval calculation, which only measures estimation uncertainty in $\hat{\theta}$. That is, $\hat{p}_1$ in (14) is set to either $\hat{p}_{1|1}$ or $\hat{p}_{1|0}$ and not varied. Once again, the boundary of the $(1-\alpha)100\%$ confidence set for $\theta_0$ is obtained by inverting the relevant Wald test in (3), with $V(\hat{\theta})^{-1}$ the relevant (numerical) estimate of the information matrix and $c_\alpha$ the appropriate critical value from the (asymptotically valid) $\chi^2(4)$ distribution. Figure 5 provides an animated plot of the full range of forecast distributions corresponding to a grid of parameter values on the surface of the 4-dimensional ellipsoid centered at $\hat{\theta}$ that is defined by (3) in this case. The settings for the illustration are $T = 100$ and $(\mu_{10}, \mu_{00}, \sigma_0^2, p_{11}, p_{10}) = (3, 0, 1, 0.6, 0.4)$, and $d_t$ is generated recursively by

$$d_t \sim Bernoulli\{s(t)\}, \quad s(t) = p_{10} + (p_{11} - p_{10}) d_{t-1}, \quad d_0 = 0, \quad t = 1, \cdots, T. \tag{15}$$

The variation in the shapes in the possible forecast distributions is marked, with bimodality a feature of many. In comparison with the empirical estimate of the forecast distribution that assigns highest probability mass to a range of values for $Y_{T+1}$ close to the origin, on this 95% boundary bi-modal distributions that assign highest probability mass to regions in the support quite *distinct* from that highlighted in the empirical forecast distribution, feature.

Figure 5: Forecast distributions at the 95th percentile boundary of the confidence set in (4), for data generated artificially from (14) and (15), with $(\mu_{10}, \mu_{00}, \sigma_0^2, p_{11}, p_{10}) = (3, 0, 1, 0.6, 0.4)$ and $T = 100$. The solid lines represent the bounding distributions, whilst the empirical distribution is given by the dotted line.

# 4 Empirical Illustration

Motivated by the increased interest in distributional forecasts of financial returns and/or volatility (see, Diebold *et al.*, 1998, Tay and Wallis, 2000, Geweke and Amisano, 2010, Maheu and McCurdy, 2011, Maneesoonthorn *et al.*, 2012, and Maheu and Jensen, 2014, for examples) in this section we consider a bivariate model for the daily return variance, $\{\mathcal{V}_t\}$, measured as (annualized) realized variance constructed from 5-minute values of the S&P500 index, and (annualized) daily returns $\{r_t\}$, where $r_t = (\log P_t - \log P_{t-1}) \times 250$, with $P_t$ the S&P500 index at the end of day $t$. The dataset spans the period September 14, 2005 to September 23, 2008.[2]

---

[2] All index data has been supplied by the Securities Industries Research Centre of Asia Pacific (SIRCA) on behalf of Reuters, with the raw index data having been cleaned using methods similar to those of Brownlees and Gallo (2006). For further details on data handing and the construction of the realized variance measure, see

We assume that $r_t$ follows the following model

$$r_t = \mu_t + \sigma_t u_t, \tag{16}$$

where $u_t \sim N(0,1)$, $\sigma_t^2 = Var_{t-1}(r_t)$ is the conditional variance of $r_t$ with respect to the information set available at time $t-1$, say $\mathcal{H}_{t-1}$, and $\mu_t = E_{t-1}(r_t) = E(r_t|\mathcal{H}_{t-1})$. We assume that

$$\mu_t = \alpha_1 + \alpha_2 r_{t-1}, \tag{17}$$

where $|\alpha_2| < 1$. In the spirit of Andersen, Bollerslev and Diebold (2007), Corsi (2009), Bollerslev, Kretschmer, Pigorsch and Tauchen (2009) and Maheu and McCurdy (2011), amongst others, we assume that $\sigma_t^2 = E_{t-1}(\mathcal{V}_t)$ and that:

$$\log(\mathcal{V}_t) = \beta_t + \varepsilon_t, \quad \varepsilon_t \sim N(0, \sigma_\mathcal{V}^2), \tag{18}$$

with:

$$\beta_t = \omega + \phi_1 \log(\mathcal{V}_{t-1}) + \phi_2 \log(\mathcal{V}_{t-5,5}) + \phi_3 \log(\mathcal{V}_{t-22,22}) + \gamma u_{t-1}, \tag{19}$$

and

$$\log(\mathcal{V}_{t-h,h}) := \frac{1}{h} \sum_{i=0}^{h-1} \log(\mathcal{V}_{t-h+i}), \quad h \geq 1, \quad (\omega, \phi_1, \phi_2, \phi_3, \gamma)' \in \mathbb{R}^5, \quad 0 < \sigma_\mathcal{V}^2 < \infty.$$

The innovation in (18), $\varepsilon_t \sim N(0, \sigma_\mathcal{V}^2)$, is assumed to be independent of that in (16), with any feedback from past shocks to volatility accommodated via the lagged innovation term in (19). The process in (19) is referred to as a heterogeneous autoregressive (HAR) model for the (log) realized variance and is designed to capture, in a simple way, the long memory that characterizes observed variance measures (see Andersen, Bollerslev, Diebold and Labys, 2003, for an early exposition of this empirical regularity). The horizons used to define the right-hand-side variables in (19) correspond to the previous day, the previous (trading) week and the previous (trading) month, with the rationale being that stock market volatility on day $t$ is the outcome of the behaviour of investors with different investment horizons.

Given $\sigma_t^2 = E_{t-1}(\mathcal{V}_t)$, it follows that $\sigma_t^2 = \exp\{E_{t-1}[\log(\mathcal{V}_t)] + 0.5 \text{Var}_{t-1}[\log(\mathcal{V}_t)]\} = \exp\{\beta_t + \sigma_\mathcal{V}^2/2\}$, and, conditional on $\mathcal{H}_{t-1}$,

$$\begin{bmatrix} r_t \\ \log(\mathcal{V}_t) \end{bmatrix} \sim N\left(\begin{bmatrix} \mu_t \\ \beta_t \end{bmatrix}, \begin{bmatrix} \exp\{\beta_t + \sigma_\mathcal{V}^2/2\} & 0 \\ 0 & \sigma_\mathcal{V}^2 \end{bmatrix}\right).$$

---

Maneesoonthorn *et al.* (2012).

Hence, the unknown parameter vector is $\theta = (\alpha_1, \alpha_2, \omega, \phi_1, \phi_2, \phi_3, \gamma, \sigma_\mathcal{V}^2)' \in \mathcal{R}^7 \times \mathcal{R}^+$ and the MLE is defined as $\hat{\theta} = (\hat{\alpha}_1, \hat{\alpha}_2, \hat{\omega}, \hat{\phi}_1, \hat{\phi}_2, \hat{\phi}_3, \hat{\gamma}, \hat{\sigma}_\mathcal{V}^2)'$.

We conduct two exercises. First, we produce animated representations of the predictive distributions on the 95% boundary for the return and its variance for a selected day at the height of the recent global financial crisis, namely September 24, 2008, in order to highlight the extent of the variation in predictive conclusions regarding both the market return and its variance that *could* have arisen - as a consequence of parameter uncertainty - in this time of extreme market volatility. Secondly, we consider an alternative (nested) bivariate model, and compute the difference in (firstly) the logarithmic scores associated with the forecast distributions (for $r_{T+1}$ and $\mathcal{V}_{T+1}$ respectively) produced by the two models - general and nested - at corresponding points in the 95% confidence set boundaries. By 'corresponding' we mean each pair of forecast distributions produced by traversing the bounding parameter space described by the corresponding ellipsoid for each of the two models, in the same 'direction'. Note that, for each of the two models, on the boundary of the conditional confidence set in (5), the plausible values for $\eta(\theta; r_{1:T}, \mathcal{V}_{1:T}) = (\mu_{T+1}, \beta_{T+1}, \sigma_\mathcal{V}^2)'$ (with the three elements of $\eta(\theta; r_{1:T}, \mathcal{V}_{1:T})$ defined according to (17), (19) and (18) respectively) describe the surface of a three-dimensional ellipsoid, which is all that is necessary to describe a confidence set for the forecast distributions of both $r_{T+1}$ and and $\mathcal{V}_{T+1}$.[3] We then produce the corresponding values for the difference in the quadratic scores for each model (and for each of the two forecast variables, $r_{T+1}$ and and $\mathcal{V}_{T+1}$). Whilst informal in nature, it of interest to document the extent to which these two different measures of relative predictive accuracy (log and quadratic score respectively) are influenced by parameter uncertainty and, in particular, whether the sign of either difference switches and the conclusion regarding predictive superiority changes, as the bounding parameter spaces are traversed.

## 4.1 Bounding sets for the distributions of $r_{T+1}$ and $\mathcal{V}_{T+1}$ via the (conditional) Wald method

Let $\eta(\theta; r_{1:t-1}, \mathcal{V}_{1:t-1}) = (\mu_t, \beta_t, \sigma_\mathcal{V}^2)'$. Then $\eta(\hat{\theta}; r_{1:t-1}, \mathcal{V}_{1:t-1}) = (\hat{\mu}_t, \hat{\beta}_t, \hat{\sigma}_\mathcal{V}^2)'$, where $\hat{\mu}_t = \hat{\alpha}_1 + \hat{\alpha}_2 r_{t-1}$ and $\hat{\beta}_t = \hat{\omega} + \hat{\phi}_1 \log(\mathcal{V}_{t-1}) + \hat{\phi}_2 \log(\mathcal{V}_{t-5,5}) + \hat{\phi}_3 \log(\mathcal{V}_{t-22,22}) + \hat{\gamma} u_{t-1}$. The relevant test statistic for $H_0 : \eta(\theta_0; r_{1:T}, \mathcal{V}_{1:T}) = \eta(\theta; r_{1:T}, \mathcal{V}_{1:T})$ against $H_1 : \eta(\theta_0; r_{1:T}, \mathcal{V}_{1:T}) \neq \eta(\theta; r_{1:T}, \mathcal{V}_{1:T})$

---

[3]The nested model contains (by construction) less base parameters than the general model. However, for both models, the confidence sets are defined by only $\eta(\theta; y_{1:T}) = (\mu_{T+1}, \beta_{T+1}, \sigma_\mathcal{V}^2)'$ and, hence, the ellipsoids that correspond to both models are defined in this same three-dimensional space and can be traversed in the same direction. More details on this point are provided in Section 4.2.

is thus given by

$$\omega\left(\eta(\theta;r_{1:T},\mathcal{V}_{1:T})\right) = T[\eta(\widehat{\theta};r_{1:T},\mathcal{V}_{1:T}) - \eta(\theta;r_{1:T},\mathcal{V}_{1:T})]'\Upsilon^{-1}[\eta(\widehat{\theta};r_{1:T},\mathcal{V}_{1:T}) - \eta(\theta;r_{1:T},\mathcal{V}_{1:T})],$$

where $\Upsilon = \nabla(\hat{\theta};r_{1:T},\mathcal{V}_{1:T})V(\hat{\theta})\nabla(\hat{\theta};r_{1:T},\mathcal{V}_{1:T})'$,

$$\begin{aligned}\nabla(\hat{\theta};r_{1:T},\mathcal{V}_{1:T}) &= \frac{\partial \eta(\theta;r_{1:T},\mathcal{V}_{1:T})}{\partial \theta'} \\ &= \begin{bmatrix} 1 & r_T & 0 & 0 & 0 & 0 & 0 & 0 \\ 0 & 0 & 1 & \log(\mathcal{V}_T) & \log(\mathcal{V}_{(T+1)-5,5}) & \log(\mathcal{V}_{(T+1)-22,22}) & u_T & 0 \\ 0 & 0 & 0 & 0 & 0 & 0 & 0 & 1 \end{bmatrix},\end{aligned}$$

and with $V(\hat{\theta})^{-1}$ being a numerical estimate of the information matrix. As noted in the previous section, on the boundary of the confidence set in (5) the plausible values for $\eta(\theta;r_{1:T},\mathcal{V}_{1:T}) = (\mu_{T+1},\beta_{T+1},\sigma_{\mathcal{V}}^2)$ describe the surface of a three-dimensional ellipsoid, which is all that is necessary to describe a confidence set for the true one-step-ahead forecast distribution for the stock index return $r_{T+1}$,

$$f_r\left(.|r_{1:T},\mathcal{V}_{1:T};\theta_0\right) \equiv N\left(\mu_{0T+1},\exp\{\beta_{0T+1}+\sigma_{0\mathcal{V}}^2/2\}\right),$$

with $\mu_{0T+1}$ and $\beta_{0T+1}$ defined according to (17) and (19) respectively, given the true values for the base parameters. The boundary of the $(1-\alpha)100\%$ confidence set for $f_r\left(\cdot|r_{1:T},\mathcal{V}_{1:T},\theta_0\right)$ is given by

$$\mathcal{B}_\alpha^r = \{f_r\left(.|r_{1:T},\mathcal{V}_{1:T};\theta\right) : \omega(\eta(\theta;r_{1:T},\mathcal{V}_{1:T})) = c_\alpha\}. \tag{20}$$

Equivalently, we can focus on the forecast distribution of the (observable) variance itself,

$$f_\mathcal{V}\left(.|r_{1:T},\mathcal{V}_{1:T};\theta_0\right) \equiv LN(\beta_{0T+1},\sigma_{0\mathcal{V}}^2),$$

and define:

$$\mathcal{B}_\alpha^\mathcal{V} = \{f_\mathcal{V}\left(.|r_{1:T},\mathcal{V}_{1:T};\theta\right) : \omega(\eta(\theta;r_{1:T},\mathcal{V}_{1:T}) = c_\alpha\} \tag{21}$$

as the boundary of the $(1-\alpha)100\%$ confidence set for $f_\mathcal{V}\left(.|r_{1:T},\mathcal{V}_{1:T};\theta_0\right)$.

In Figure 6, we provide an animated representation of the full range of bounding distributions for $f_r\left(.|r_{1:T},\mathcal{V}_{1:T};\theta_0\right)$ produced from the boundary set in (20), with $1-\alpha = 0.95$. The corresponding animated graph of the bounding distributions for $f_\mathcal{V}\left(.|r_{1:T},\mathcal{V}_{1:T};\theta_0\right)$ defined by (21) are displayed in Figure 7. The results are based on $T = 762$ observations for the period September 14, 2005 to September 23, 2008 leading up to the date September 24, 2008, for which predictions are made, and on which values of $r_{T+1} = -0.65$ and $\mathcal{V}_{T+1} = 0.064$ were observed. Once again, the graphs in both figures indicate the range of different possible outcomes that *could* be

Figure 6: Forecast distributions at the 95th percentile boundary of the confidence set in (20). The results are conditional on $T = 762$ observations for the period September 14, 2005 to September 23, 2008 leading up to the date September 24, 2008, for which predictions are made. The solid line represents the bounding distributions. The dotted line depicts the estimated distribution. The vertical solid line represents the observed value of $r_{T+1}$, which is $-0.65$.

Figure 7: Forecast distributions at the 95th percentile boundary of the confidence set in (21). The results are conditional on $T = 762$ observations for the period September 14, 2005 to September 23, 2008 leading up to the date September 24, 2008, for which predictions are made. The solid line represents the bounding distributions. The dotted line depicts the estimated distribution. The vertical solid line represents the observed value of $\mathcal{V}_{T+1}$, which is $0.064$.

observed in hypothetical repeated sampling and, hence, the type of variation to be expected in any probabilistic statements made about $r_{T+1}$ (Figure 6) and $\mathcal{V}_{T+1}$ (Figure 7), respectively.

In particular, first with reference to the return (in Figure 6), whilst the empirical forecast distribution assigns a very low density value to the observed (annualized) return on September 24, namely $-0.65$, the influence of parameter variation is such that a negative return of this magnitude - and the consequences of that for any associated financial decisions - *could* have been assigned *either* a much lower *or* a much higher density value! Corresponding to this variation in possible outcomes, predictions of the one-day-ahead 5% Value at Risk (VaR) quantile for the market portfolio associated with the S&P500 index would have produced either a notable violation (i.e. the observed portfolio value being much *less* than the VaR value) or a clear absence of such violation. Extrapolating these consequences to a realistic setting in which portfolios are designed to track the market, and financial penalties are incurred for repeat violations of (or overly conservative) VaR forecasts, parameter variation - and accommodation thereof - is seen to have clear practical significance.

In contrast, the influence of parameter variation on the forecast distribution for the variance itself (in Figure 7) is seen to be much less extreme, with the observed value of 0.064 (equivalent to an annualized standard deviation (volatility) of approximately 25%) assigned a density ordinate by the empirical forecast distribution that varies little from those associated with the extreme 95% boundary. This could be interpreted as a certain robustness of the variance forecast - and any contingent financial decisions, such as derivative pricing - to parameter variation, at least conditional on the given model. In the following section we go one step further, and assess the robustness (or otherwise) of the difference in scores - for two different models - to parameter uncertainty.

## 4.2 Bounding values for scoring rule differences

Scoring rules are scalar measures used to assess the relative performance of competing probabilistic forecasts. In this illustration, we consider two proper scoring rules: the logarithmic score (LS) and the quadratic score (QS), given respectively (and for the case of predicting the return, for illustration) by

$$LS = \log f_r(r_{T+1}^o|r_{1:T}, \mathcal{V}_{1:T}; \theta_0) \tag{22}$$

$$QS = 2f_r(r_{T+1}^o|r_{1:T}, \mathcal{V}_{1:T}; \theta_0) - \int [f_r(r_{T+1}|r_{1:T}, \mathcal{V}_{1:T}; \theta_0)]^2 dr_{T+1}, \tag{23}$$

where $r_{T+1}^o$ denotes the observed value of $r_{T+1}$. The $LS$ in (22) is a so-called 'local' scoring rule which assumes a high value if $r_{T+1}^o$ is in the high density region of $f_r(.|r_{1:T}, \mathcal{V}_{1:T}; \theta_0)$ and a low value otherwise. In contrast, $QS$ depends on the shape of the entire predictive density, in addition to the ordinate of the density at the realized value of $r_{T+1}$. In particular, $QS$ combines a reward for a well-calibrated prediction (a high value of $f_r(r_{T+1}^o|r_{1:T}, \mathcal{V}_{1:T}; \theta_0)$) with a penalty $(-\int [f_r(r_{T+1}|r_{1:T}, \mathcal{V}_{1:T}; \theta_0)]^2 dr_{T+1})$ for misplaced 'sharpness', or certainty, in the prediction. That is, if $f_r(r_{T+1}|r_{1:T}, \mathcal{V}_{1:T}; \theta_0)$ is a concentrated density (and not necessarily around $r_{T+1}^o$), this penalty will be high. (See Gneiting and Raftery, 2007, Gneiting, Balabdaoui and Raftery, 2007, and Boero, Smith and Wallis, 2011, for expositions). Comparable definitions and interpretations apply to the $LS$ and $QS$ scoring rules computed for $\mathcal{V}_{T+1}$.

Denoting the model given by (16) to (19) by $M_1$, we now consider the following nested version, denoted by $M_2$, in which a short memory structure for $\mathcal{V}_t$ is imposed:

$$r_t = \mu_t + \sigma_t u_t, \quad \mu_t = \alpha_1 + \alpha_2 r_{t-1}, \quad u_t \sim N(0,1), \quad |\alpha_2| < 1, \tag{24}$$

$$\sigma_t^2 = Var_{t-1}(r_t) = E_{t-1}(\mathcal{V}_t), \quad \log(\mathcal{V}_t) = \beta_t + \varepsilon_t, \quad \varepsilon_t \sim N(0, \sigma_\mathcal{V}^2), \tag{25}$$

$$\beta_t = \omega + \phi_1 \log(\mathcal{V}_{t-1}) + \gamma u_{t-1}, \tag{26}$$

$$(\alpha_1, \alpha_2, \omega, \phi_1, \gamma)' \in \mathbb{R}^5, \ 0 < \sigma_\mathcal{V}^2 < \infty, \tag{27}$$

Whilst we would not necessarily expect $M_2$ to provide more accurate probabilistic forecasts than model $M_1$, given the more restricted dependence structure, it is of interest to see if the expected superiority of the long memory model *does* obtain for the short horizon and if it is uniform over the bounding parameter space. To this end, we first provide, in Figure 8, an animated representation of the full range of bounding distributions for $r_{T+1}$ for both models: $M_1$ and $M_2$. The corresponding animated graphs of the bounding pdfs for $\mathcal{V}_{T+1}$ under the two models are displayed in Figure 9.

With regard to the two forecast densities for $r_{T+1}$ there is clearly a *substantial* difference between the impact of parameter variation on the density under $M_1$ and the corresponding impact under $M_2$, with the former density changing markedly in location and, hence, producing very different values for the ordinate at the observed value of $r_{T+1} = -0.65$. This, in turn, implies that the relative magnitudes of the log scores for the two models also change markedly at the 95% boundary. As a consequence, in Panel A of Figure 10 the difference in log scores is shown to be either positive or negative depending on the precise position on the two bounding ellipsoids. That is, despite the fact that the difference in empirical scores (not displayed) is positive - i.e. $M_1$ is estimated to have better predictive performance than $M_2$ - the conclusion that one *could* have drawn as to which model were superior switches due to sampling variation! A qualitatively similar story - although with a somewhat different pattern exhibited for the log score differences - holds for prediction of $\mathcal{V}_{T+1}$ (illustrated in Panel C).

Figure 8: Forecast distributions at the 95th percentile boundary of the confidence set in (20), for the Model $M_1$ (solid line) and Model $M_2$ (dashed line). The setting is the same as that for Figure 6. The vertical solid line represents the observed value of $r_{T+1}$ which is $-0.65$.

Figure 9: Forecast distributions at the 95th percentile boundary of the confidence set in (21), for the Model $M_1$ (solid line) and Model $M_2$ (dashed line). The setting is the same as that for Figure 7. The vertical solid line represents the observed value of $\mathcal{V}_{T+1}$, which is 0.064.

In contrast, if the difference in the quadratic scores is used to assess relative performance, parameter variation is seen (Panels B and D in Figure 10) to have *no* qualitative impact, with the more general model remaining superior for all points on the bounding ellipsoids, and for the prediction of both random variables. In this sense the quadratic score could be deemed to be more robust to parameter variation than the log score, at least as concerns the comparison of these two particular models.

We conclude by noting that the precise patterns exhibited by all graphs in Figure 10 across the 441 grid points that cover the bounding ellipsoids for the two different models simply reflect the particular *order* in which those grid points are used to evaluate the forecast distributions (and the resultant impact on the relevant scores).[4] The key thing is that a common parameter space, $\eta(\theta; r_{1:T}, V_{1:T}) = (\mu_{T+1}, \beta_{T+1}, \sigma_{\mathcal{V}}^2)$, characterizes both models and that the common space is traversed in the same direction for both models. Had a different common direction been taken, then the patterns exhibited would be different; however, the relative values (at any given grid point) of the two scores - logarithmic or quadratic - would be the same as those displayed in Figure 10, those relative values reflecting the values of $(\mu_{T+1}, \beta_{T+1}, \sigma_{\mathcal{V}}^2)'$ for each model, at the particular grid point.

For illustration of this point, we let $(\mu_{T+1}^{(1)}, \beta_{T+1}^{(1)}, \sigma_{\mathcal{V}}^{2(1)})$ and $(\mu_{T+1}^{(2)}, \beta_{T+1}^{(2)}, \sigma_{\mathcal{V}}^{2(2)})$ denote the boundary values of $(\mu_{T+1}, \beta_{T+1}, \sigma_{\mathcal{V}}^2)$, for $M_1$ and $M_2$ respectively, represented, in turn, by grid points on the surface of the two ellipsoids presented in Figure 11. It can be shown that the difference in log scores (for forecasting $r_{T+1}$) has the following closed-form representation,

$$
\begin{aligned}
LS_1 - LS_2 &= \frac{1}{2}\left(\beta_{T+1}^{(2)} - \beta_{T+1}^{(1)}\right) + \frac{1}{4}\left(\sigma_{\mathcal{V}}^{2(2)} - \sigma_{\mathcal{V}}^{2(1)}\right) \\
&+ \frac{1}{2}\left(r_{T+1} - \mu_{T+1}^{(2)}\right)^2 \exp\left\{-(\beta_{T+1}^{(2)} + \sigma_{\mathcal{V}}^{2(2)}/2)\right\} \\
&- \frac{1}{2}\left(r_{T+1} - \mu_{T+1}^{(1)}\right)^2 \exp\left\{-(\beta_{T+1}^{(1)} + \sigma_{\mathcal{V}}^{2(1)}/2)\right\},
\end{aligned} \quad (28)
$$

where the different terms on the right-hand-side of (28) capture the influence of the various boundary values on $LS_1 - LS_2$ and, hence, explain the reason for the types of values plotted in Panel A of Figure 10.

---

[4]The grid points are generated by using the Matlab function '*ellipsoid*'. In Matlab, the function $[x, y, z] = ellipsoid(xc, yc, zc, xr, yr, zr, n)$ generates a surface mesh described by three $n + 1$-by-$n + 1$ matrices. These $(x, y, z)$ coordinates describe a surface of an ellipsoid with center $(xc, yc, zc)$ and semi-axis lengths $(xr, yr, zr)$. We used $n = 20$ and hence obtained 441 grid points.

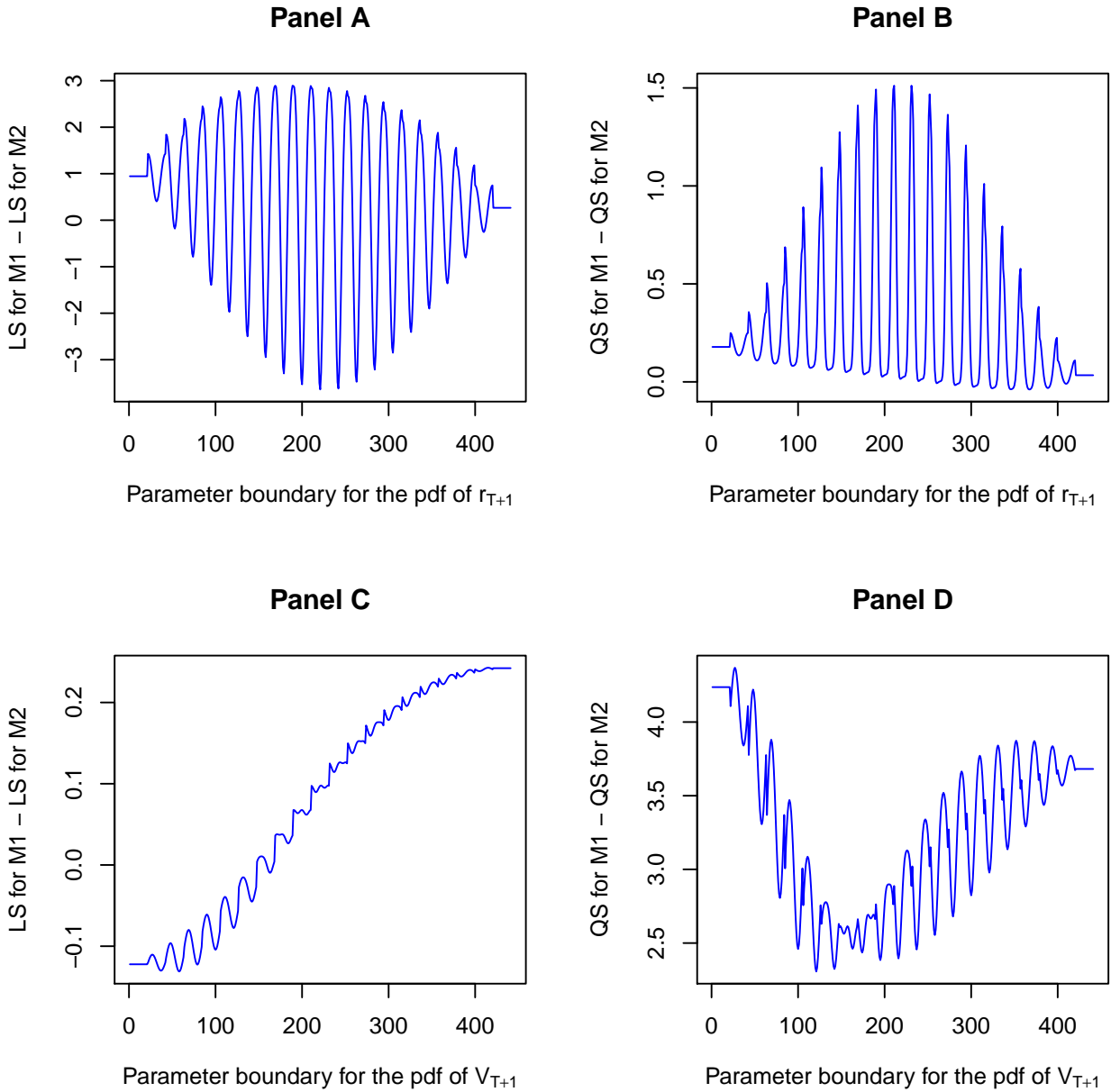

Figure 10: Panel A [Panel B]: Differences in LS [QS] scores that correspond to the bounding distributions for $r_{T+1}$, for the models $M_1$ and $M_2$. Panel C [Panel D]: Differences in LS [QS] scores that correspond to the bounding distributions for $\mathcal{V}_{T+1}$, for the models $M_1$ and $M_2$. The results are conditional on $T = 762$ observations for the period September 14, 2005 to September 23, 2008, leading up to the date September 24, 2008, for which predictions are made.

Figure 11: Panel A: Ellipsoid defined by the 95th percentile boundary of the confidence set in (5) for model $M_1$ (general model). Panel B: Ellipsoid defined by 95th percentile boundary of the confidence set in (5) for model $M_2$ (nested model). The results are conditional on $T = 762$ observations of intraday spot price data from the S&P500 index for the period September 14, 2005 to September 23, 2008 prior to the date, September 24, 2008, for which an estimated forecast distribution is produced.

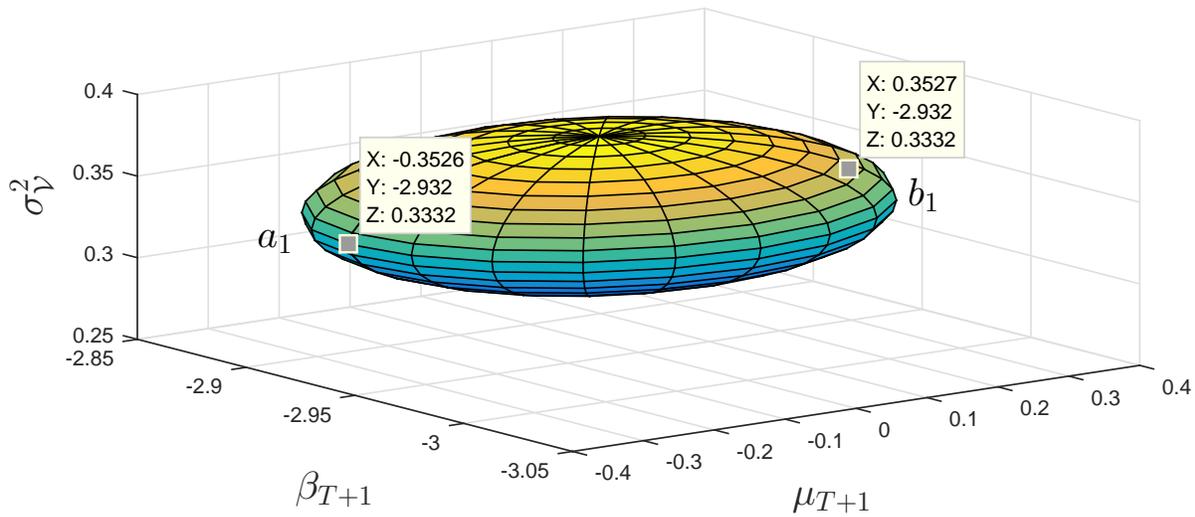

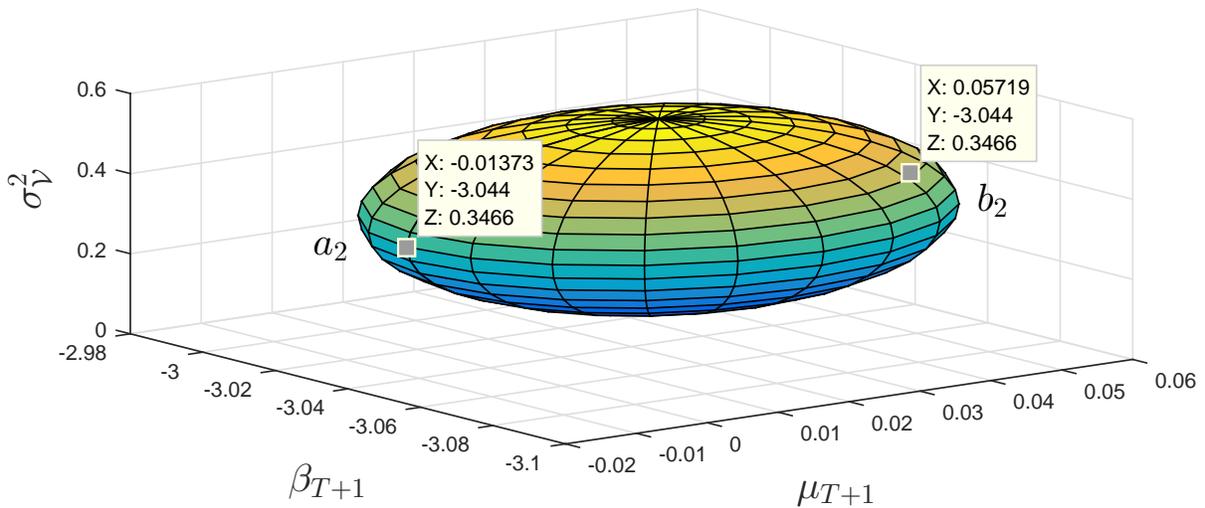

With reference to the ellipsoid in Panel A of Figure 11 (for $M_1$), $a_1 = (\mu_{T+1}^{(1)}, \beta_{T+1}^{(1)}, \sigma_{1\mathcal{V}}^2) = (-0.3526, -2.932, 0.3332)$ and $b_1 = (\mu_{T+1}^{(1)}, \beta_{T+1}^{(1)}, \sigma_{1\mathcal{V}}^2) = (0.3527, -2.932, 0.3332)$ represent the two points, among all the values on the surface mesh, that are furthest apart along the $X$-axis (i.e. $\mu_{T+1}^{(1)}$ values). Points $a_2 = (\mu_{T+1}^{(2)}, \beta_{T+1}^{(2)}, \sigma_{2\mathcal{V}}^2) = (-0.0137, -3.044, 0.3466)$ and $b_2 = (\mu_{T+1}^{(2)}, \beta_{T+1}^{(2)}, \sigma_{2\mathcal{V}}^2) = (0.0572, -3.044, 0.3466)$ denote the corresponding pair of points on the ellipsoid in Panel B of Figure 11 (for $M_2$). For points $a_1$ and $a_2$, it can be seen that the associated pairs of values, $(\beta_{T+1}^{(1)}, \sigma_{\mathcal{V}}^{2(1)})$ and $(\beta_{T+1}^{(2)}, \sigma_{\mathcal{V}}^{2(2)})$, are quite similar, such that the first two terms on the right-hand-side of (28) are very small, and the exponential terms that feature in the next two terms are very similar in magnitude. In contrast, the two values $\mu_{T+1}^{(1)}$ and $\mu_{T+1}^{(2)}$ are quite different, one from the other. Given that $r_{T+1} = -0.65$, and since $\mu_{T+1}^{(1)} = -0.3526 < -0.0137 = \mu_{T+1}^{(2)}$, it follows that $\left(r_{T+1} - \mu_{T+1}^{(2)}\right)^2 > \left(r_{T+1} - \mu_{T+1}^{(1)}\right)^2$, and that $LS_1 - LS_2 > 0$ as a consequence. This $LS_1 - LS_2$ value corresponds to the largest positive peak of the graph in Panel A of Figure 10 (occurring at grid point 211). Similarly, for the two points $b_1$ and $b_2$ on the relevant ellipsoids, $\mu_{T+1}^{(1)} = 0.3527 > 0.0572 = \mu_{T+1}^{(2)}$ and values (for the two models) for $\beta_{T+1}$ and $\sigma_{\mathcal{V}}^2$ are once again similar. In this instance then, $LS_1 - LS_2 < 0$, yielding the largest negative value of the graph in Panel A of Figure 10 (occurring at grid point 221). The cyclic nature of $LS_1 - LS_2$ in Panel A of Figure 10 is simply a reflection of traversing the two ellipsoids in this fashion, from points such as $(a_1, a_2)$ to points such as $(b_1, b_2)$. Comparable explanations can be provided for the patterns exhibited in the remaining panels in Figure 10.

## 5 Discussion

We have used the inversion of an optimal test to produce a confidence region for a distributional forecast that is asymptotically uniformly most accurate. The method moves away both from the idea of placing point-wise confidence intervals on the ordinates of the forecast density (or mass) function, and from the literature's typical focus on bootstrap prediction intervals/regions. The method is also completely general, and simple to implement, with techniques of dimension reduction available in many cases via conditioning. Visualization of the bounding distributions is possible using animated graphics, enabling the full range of distributions that *could* have arisen as a result of parameter uncertainly to be clearly displayed. Whilst the 95th percentile of the sampling distribution (for the estimated forecast distribution) has been used throughout for illustration, a comparable type of computation (and display) could of course be undertaken for *any* level of confidence.

Documenting the impact of sampling variation on relevant scalar functions, such as scoring rules, is straightforward and has been illustrated in an empirical setting via the computation of the log and quadratic scores. It has been possible to match the bounding ellipsoids in the case of the nested models under comparison here and, hence, provide unambiguous results regarding the behaviour of the score difference as the two 95% boundaries are traversed. In contrast, for *non-nested* models, there is not necessarily a unique way in which forecast distributions for two competing models can be 'matched' and differences in the log scores computed, and the resolution of this ambiguity remains to be addressed.

Finally, as fits with convention, parameter uncertainty only has been the focus, with the conditioning values viewed as fixed numbers. Any move away from this approach would require the confidence set to accommodate the joint distribution of the estimated parameters and the conditioning values; something that is the subject of ongoing work by the authors.

# References

[1] Amisano, G. and Giacomini, R. 2007. Comparing Density Forecasts via Weighted Likelihood Ratio Tests, *Journal of Business and Economic Statistics* 25, 177-190.

[2] Andersen, T.G., Bollerslev, T., Diebold, F.X. and Labys, P. 2003. Modelling and Forecasting Realized Volatility, *Econometrica* 71, 579-625.

[3] Andersen, T.G., Bollerslev, T. and Diebold, F.X. 2007. Roughing It Up: Including Jump Components in the Measurement, Modeling and Forecasting of Return Volatility, *The Review of Economics and Statistics* 89, 701-720.

[4] Boero, G., Smith, J. and Wallis, K.F. 2011. Scoring Rules and Survey Density Forecasts. *International Journal of Forecasting* 27, 379-393.

[5] Bollerslev, T., Kretschmer, U., Pigorsch, C. and Tauchen, G. 2009. A Discrete-Time Model for Daily S&P500 Returns and Realized Variations: Jumps and Leverage Effects, *Journal of Econometrics* 150, 151-166.

[6] Brownlees, C.T. and Gallo, G.M. 2006. Financial Econometric Analysis at Ultra-High Frequency: Data Handling Concerns, *Computational Statistics and Data Analysis* 51, 2232-2245.

[7] Choi, S., Hall, W.J. and Schick, A. 1996. Asymptotically Uniformly Most Powerful Tests in Parametric and Semiparametric Models, *The Annals of Statistics* 24, 841-861

[8] Corradi, V. and Swanson, N. 2006. Predictive Density and Conditional Confidence Interval Accuracy Tests, *Journal of Econometrics* 135, 187-228.

[9] Corsi, F. 2009. A Simple Approximate Long Memory Model of Realized Volatility, *Journal of Financial Econometrics* 7, 174-196.

[10] Cox, D.R. and Hinkley, D.V. 1974. *Theoretical Statistics*, Chapman and Hall, London.

[11] Czado, C., Gneiting, T. and Held, L. 2009. Predictive Model Assessment for Count Data, *Biometrics* 65, 1254-1261

[12] Dawid, A.P. 1984. Present Position and Potential Developments: Some Personal Views. Statistical Theory. The Prequential Approach, *Journal of the Royal Statistical Society (A)* 147, 278–292.

[13] De Gooijer, J.G. and Hyndman, R. 2006. 25 Years of Time Series Forecasting, *International Journal of Forecasting*, 22, 443-473.

[14] Diebold, F.X., Gunther, T.A. and Tay, A.S. 1998. Evaluating Density Forecasts with Applications to Financial Risk Management. *International Economic Review* 39, 863-883.

[15] Freeland, R and McCabe, B. 2004. Forecasting Discrete Valued Low Count Time Series. *International Journal of Forecasting* 20, 427-434.

[16] Geweke, J. and Amisano, G. 2010. Comparing and Evaluating Bayesian Predictive Distributions of Asset Returns, Special Issue on Bayesian Forecasting in Economics, *International Journal of Forecasting*, 26, 216-230.

[17] Gneiting, T. and Raftery, A.E. 2007. Strictly Proper Scoring Rules, Prediction, and Estimation. *Journal of the American Statistical Association* 102, 359-378.

[18] Gneiting, T., Balabdaoui, F. and Raftery, A. 2007. Probabilistic Forecasts, Calibration and Sharpness, *Journal of the Royal Statistical Society (B)* 69, 243-268.

[19] Gneiting, T. and Katzfuss, M. 2014. Probabilistic Forecasting, *Annual Review of Statistics and Its Application* 1, 125-151.

[20] Geweke, J. 2005. *Contemporary Bayesian Econometrics and Statistics,* Wiley.

[21] Jensen, M.J. and Maheu, J.M. 2014. Estimating a Semiparametric Asymmetric Stochastic Volatility Model with a Dirichlet Process Mixture, *Journal of Econometrics* 178, 523-538

[22] Le Cam, L. and Yang, G.L. 1990. *Asymptotics in Statistics: Some Basic Concepts.* Springer, New York.

[23] McCabe, B., Martin, G.M. and Harris, D. 2011. Efficient Probabilistic Forecasts for Counts. *Journal of the Royal Statistical Society (B)* 73, 253-272.

[24] Maheu, J.M. and McCurdy, T.H. 2011. Do High-Frequency Measures of Volatility Improve Forecasts of Return Distributions? *Journal of Econometrics* 160, 69-76.

[25] Maneesoonthorn, W., Martin, G.M., Forbes, C.S. and Grose, S.G. 2012. Probabilistic Predictions of Volatility and its Risk Premia, *Journal of Econometrics* 171, 217-236.

[26] Phillips., P. C. B. 1979. The Sampling Distribution of Forecasts from a First-Order Autoregression, *Journal of Econometrics* 9, 241-261.

[27] Politis, D.N., Romano, J.P. and Wolf, M. 1999. *Subsampling,* Springer, New York.

[28] Rodriguez, A. and Ruiz, E. 2009. Bootstrap Prediction Intervals in State-Space Models. *Journal of Time Series Analysis* 30, 167-178.

[29] Rodriguez, A. and Ruiz, E. 2012. Bootstrap Prediction Mean Squared Errors of Unobserved States Based on the Kalman Filter with Estimated Parameters. *Computational Statistics and Data Analysis* 56, 62-74.

[30] Shao, J. 2003. *Mathematical Statistics*, Springer-Verlag, New York.

[31] Tay, A.S. and Wallis, K. 2000. Density Forecasting: A Survey, *Journal of Forecasting*, 19, 235-254.

[32] Wald, A. 1943. Tests of Statistical Hypotheses Concerning Several Parameters When the Number of Observations is Large, *Trans. Amer. Math. Soc.* 54, 426-482.

[33] Wolf, M. and Wunderli, D. 2015. Bootstrap Joint Prediction Regions, *Journal of Time Series Analysis*, 36, 352-376.